\documentclass[10pt,journal]{IEEEtran}
\usepackage{amsrefs}
\newenvironment{rezabib}
  {\bibdiv\biblist\setupbib}
  {\endbiblist\endbibdiv}

\def\setupbib{\catcode`@=\active}
\begingroup\lccode`~=`@
  \lowercase{\endgroup\def~}#1#{\gatherkey{#1}}
\def\gatherkey#1#2{\gatherkeyaux{#1}#2\gatherkeyaux}
\def\gatherkeyaux#1#2,#3\gatherkeyaux{\bib{#2}{#1}{#3}}

\usepackage{ulem}
\usepackage{times}
\usepackage{epsfig}
\usepackage{graphicx}
\usepackage{amsmath}
\usepackage{amssymb}
\usepackage{amsmath}
\DeclareMathOperator*{\argmin}{argmin}

\title{Residual Switching Network for Portfolio Optimization}

\author{ \textbf{Jifei Wang, Lingjing Wang}\\ 
NYU Courant Institute of Mathematical Sciences\\ 
\{jifei.wang, lingjing.wang\}@courant.nyu.edu 
}

\begin{document}

\maketitle

\begin{abstract}
This paper studies deep learning methodologies for portfolio optimization in the US equities market. We present a novel residual switching network that can automatically sense changes in market regimes and switch between momentum and reversal predictors accordingly. The residual switching network architecture combines two separate residual networks (ResNets), namely a switching module that learns stock market conditions, and the main module that learns momentum and reversal predictors. We demonstrate that over-fitting noisy financial data can be controlled with stacked residual blocks and further incorporating the attention mechanism can enhance powerful predictive properties. Over the period 2008 to H12017, the residual switching network (Switching-ResNet) strategy verified superior out-of-sample performance with an average annual Sharpe ratio of 2.22, compared with an average annual Sharpe ratio of 0.81 for the ANN-based strategy and 0.69 for the linear model.\\

\end{abstract}

\section{Introduction}

Deep learning has been extensively applied to large financial datasets in recent years. Researchers have extended the application of ANN as a non-parametric approach to optimize portfolios and built algorithmic trading strategies that are orthogonal to conventional statistical arbitrage strategies \cite{avellaneda2010statistical,dixon2018sequence}. It began with shallower machine learning models such as one-layer artificial neural networks (ANN) \cite{gately1995neural} and support vector machines \cite{cherkassky2004practical}, which demonstrated powerful predictive properties on financial time series \cite{huang2005forecasting,das2012support,dixon2018high}. As time progressed, the research community ventured into deeper non-linear topological structures for financial signal representation \cite{deng2017deep,bao2017deep} and observed significant improvement over the performance of traditional machine learning methods \cite{simonyan2014very,szegedy2015going}. \\

\begin{figure}[h]
\begin{center}
\includegraphics[width=8.5cm]{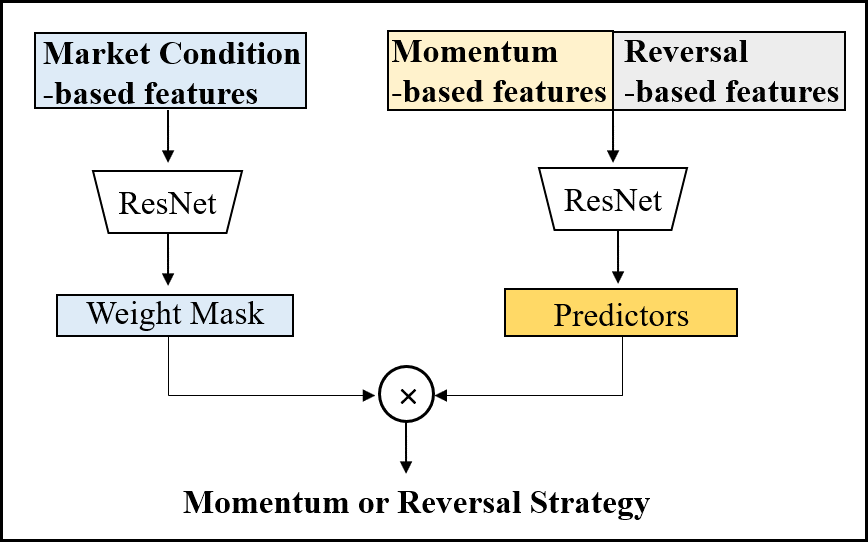}
\end{center}
\caption{Switching-ResNet combines the regime-switching module (left) with the main module (right). The conditional weight mask combines with predictors by element-wise product, $\otimes$. As market conditions change over time, our network can switch between momentum-based and reversal-based strategies accordingly.}
\label{first}
\end{figure}

The application of deep learning in finance can be broadly categorized by three approaches: stacked auto-encoders (SAE) \cite{bao2017deep,takeuchi2013applying}, recurrent neural networks (RNN) \cite{deng2017deep,bao2017deep,saad1998comparative}, and feedforward neural networks (FNN) \cite{kaastra1996designing}. Bao et al. used stacked auto-encoders to hierarchically extract deep financial features in an unsupervised manner, the features are then passed through the LSTM, a type of RNN with feedback links attached to certain layers of the network, to predict future trends in the stock market. Many researchers have extensively studied the use of feed-forward network for financial asset predictions \cite{saad1998comparative,dixon2015implementing}, and Deng et al. successfully applied RNN to make high frequency trading decisions whilst minimizing transaction cost and slippage. 

It is challenging to fit ``substantially deep" networks on noisy non-stationary financial data. Even though non-linear relationships exist in financial markets, forcefully fitting an extremely non-linear model such as a 22-layer plain ANN, often results in the tendency to over-fit, causing out-of-sample accuracy/performance to deteriorate. When fitting ``substantially deep" plain networks on stock price data, regularization methods such as dropout and weight penalization are difficult to control.\\

In this paper, we first address the over-fitting problem by applying Attention ResNet to noisy financial data to predict US equities' monthly returns. We evaluate and compare the in-sample and out-of-sample performance of the deep Attention ResNet with a regularized deep plain ANN, as well as the linear model. We demonstrate that stacked attention enhanced residual blocks can strike a balance between over-fitting (deep plain ANN) and under-fitting (linear model), and optimize the degree of non-linearity when modeling stock price fluctuations. Our proposed architecture also consists of attention masks that interface with hidden layers to generate attention enhanced financial feature representations. \\

\begin{figure*}[h]
\centering
\includegraphics[width=12cm]{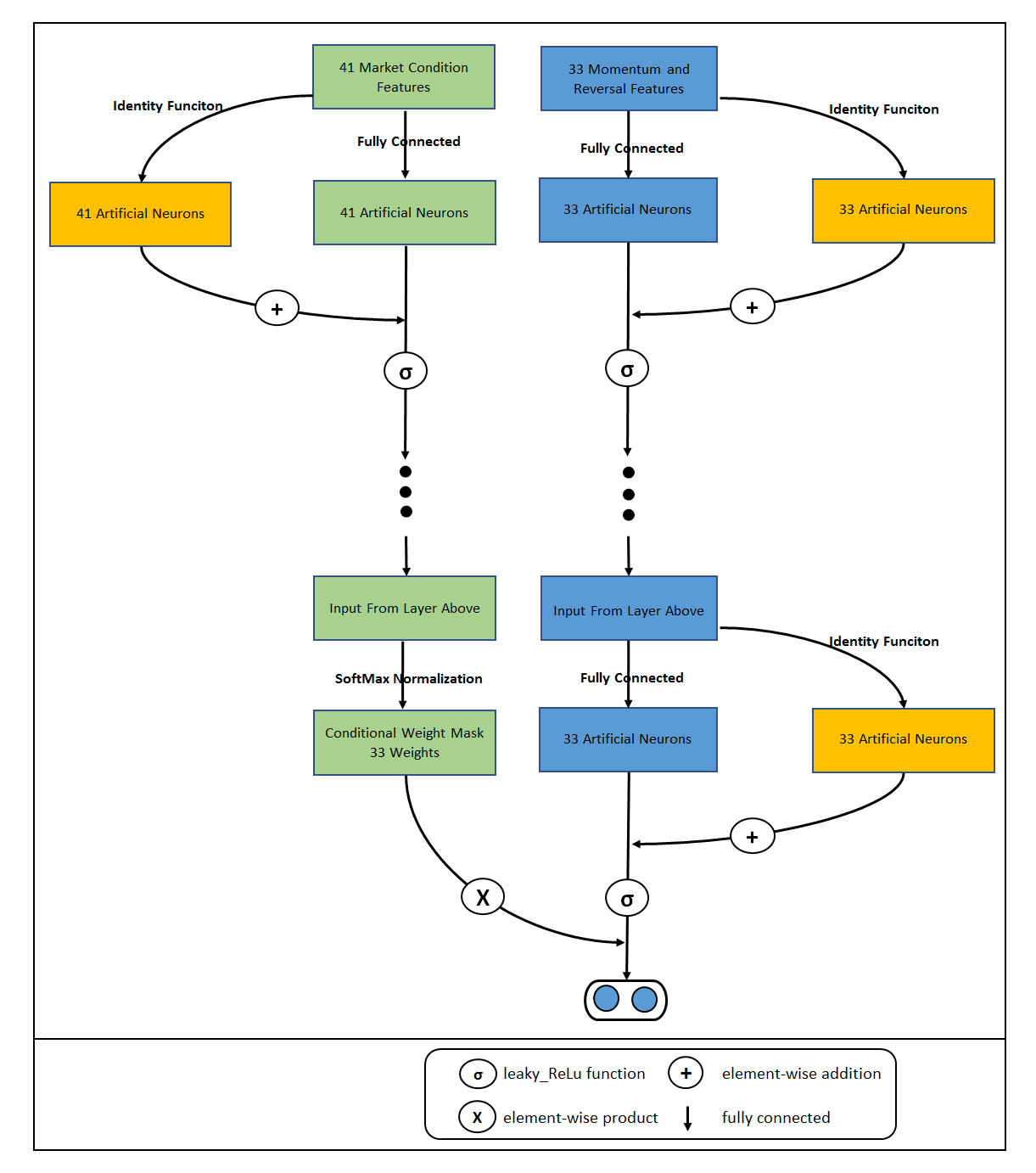}
\caption{Switching-ResNet combines the main module (ResNet in blue) with the switching module (ResNet in green). The main module learns momentum and reversal predictors of 2000 individual stocks. The switching module learns market conditions of SP500 realized volatility, squared VIX and SP500 variance risk premium. The switching module outputs a conditional weight mask, which is combined with the last parameter layer of the main ResNet (via element-wise product) to select long and short portfolios.}
\label{main}
\end{figure*}

We then present the residual switching network that automatically senses a change in financial market conditions and switch between prominent market anomalies accordingly. There are various stylized predictors in portfolio management, the two most prominent market anomalies are momentum and reversal \cite{vayanos2013institutional}. We develop a robust switching module that weighs the relative importance of momentum and reversal anomalies conditional on SP500 realized volatility, squared VIX and SP500 variance risk premium, see Fig \ref{first} and \ref{main}. Our method is based on the notion that market anomalies are directly associated with market regimes, where a bullish market regime is often associated with lower market volatility, and the bearish market regime is associated with higher market volatility \cite{schwert1989does}. As an intermediate result, we demonstrate that the standalone switching module can accurately model and predict SP500 realized volatility. We then combine the switching module with a residual network to complete the Switching ResNet. The switching module computes a conditional weight mask, which is applied to the last hidden layer of the main ResNet to identify relative important stylized predictors (momentum or reversal hidden units) as market regime changes over time. Finally, the switching module is combined with the main ResNet to select long/short portfolio (delta neutral) to reap higher rewards in portfolio management. The main contributions of our proposed method are listed as follows:   \\

\begin{itemize}

\item We introduce a novel supervised learning approach for portfolio management in the US equities market. The Attention-ResNet significantly increases network depth for applying deep learning methods to noisy financial data, whilst finding an optimal balance between over-fitting and under-fitting.\\

\item We leverage the attention module to guide abstract financial feature learning to improve predictive accuracy as well as model flexibility. \\

\item We introduce a novel Switching-ResNet that can learn financial market conditions to guide our proposed model to switch between momentum and reversal predictors as market regime changes, which significantly differentiates the Switching-ResNet from previous learning-based approaches for portfolio management.    \\

\end{itemize}

\bigskip

\section{Related Works}

Recent advancements in residual learning allow researchers to exploit much deeper network architectures than its predecessors \cite{he2016deep}. Residual learning has been extensively applied in computer vision and has proven to be a successful supervised learning approach. Residual network (ResNet) is constructed by stacking residual blocks, which are modified from plain ANNs using short circuits so that it is easier for ResNets to learn the identity mapping rather than an un-referenced mapping \cite{he2016deep}. When modeling noisy financial data, the ResNet can increase network depth whilst control for over-fitting. \\

Attention module enhances feature representations at selected focal points \cite{itti2001computational,walther2002attentional}. Previous works have shown that attention module can be easily applied to the feed-forward network architecture, and the incorporation of stacked attention modules can capture both low and high-level refined features that lead to consistent performance improvement for image classification and machine translation tasks \cite{bahdanau2014neural}. By contrast, our proposed method demonstrates that the concept behind residual learning and attention module is cross-fertilizing and hopeful for model-driven portfolio optimization.\\

Hamilton (1989) first introduced the Markov regime-switching model. This auto-regressive regime-switching process involves multiple equations to characterize multiple regimes, where the switching module is regulated by past values of an unobserved state variable \cite{hamilton1989new}. The Markov switching model and its many variants have been extensively applied to economic and financial time series. Kim and Nelson’s regime-switching model examines business cycle duration dependency to see whether a regime change depends on how long an economy was in a boom or recession \cite{kim1998business}. Researchers also experimented with variants of the Markov switching model in the foreign exchange market and found supporting evidence for its superior predictive power on the direction of change for exchange rates \cite{engel1994can}. Inspired by Markov regime-switching model and recent advances in attention mechanism for deep learning, we are the first to incorporate a regime-switching module on the deep residual network. Our proposed deep learning framework enhances portfolio management by allowing the switching module to switch between prominent stock market anomalies when market condition changes over time.\\

\section{Our Approach}
In this section, we introduce the Attention-ResNet and the Switching-ResNet. We first present the self-attention concept to establish the self-attention enhanced residual block and constructs the 22-layer deep Attention-ResNet in section 3.1. Section 3.2 discusses the details of the switching module and analyze the guided-attention concept behind the switching module. We then construct the deep residual switching network. The model configurations and training procedures are described in section 3.3.

\subsection{3.1 Attention Enhanced ResNet (Self-Attention)}\label{sc_def_net}

\begin{figure}[h]
\begin{center}
\includegraphics[width=8.5cm]{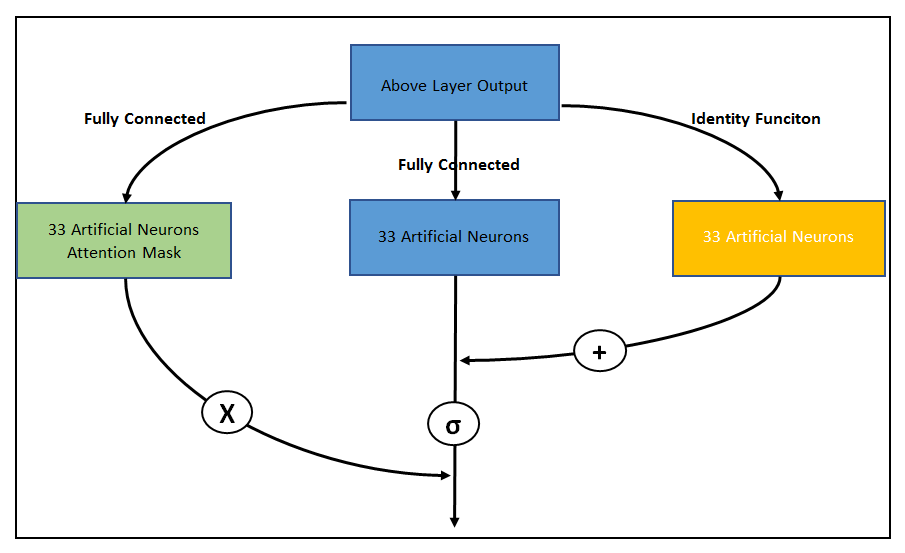}
\end{center}
\caption{Attention Enhanced Residual Block, $\oplus$ denotes element-wise addition, $\sigma$ denotes Leaky-ReLU activation function, and $\otimes$ denotes element-wise product. The short circuit in yellow occurs before $\sigma$ activation, and attention mask in green is applied after $\sigma$ activation.}
\label{fig3}
\end{figure}

The self-attention mechanism can be naturally combined with residual block to guide financial feature learning \cite{wang2017residual}. The self-attention mask estimate a set of weights learned from the input to residual block using a fully connected layer followed by SoftMax function to normalize weights to a range from 0 to 1. Shown in figure~\ref{fig3}, the attention mask weights are applied to the output units of the residual block via element-wise product (i.e. after the $\sigma$ activation function). Formally, the self-attention enhanced residual block is formulated as follows: 

Given $n \in \big\{1,2,...,N\big\}$, $i,j \in \big\{1,2,3,...,D\big\}$, ${f}^{(0)}(X)=X$ is the input vector. In (1) and (2), the attention module learns a hidden layer on ${f}^{(n-1)}(X)$ the input vector to the block. Equations (3) and (4) computes the mask weights ${f}^{a,(n+1)}(X)$, which normalizes ${z}^{a,(n+1)}(X)$ via SoftMax function, $\Phi$. In (5), we apply attention mask weights ${f}^{a,(n+1)}(X)$ to the output of residual block, ${f}^{(n+1)}(X)$ , via element-wise product.\\

Self-Attention Module,
\begin{eqnarray}
\label{eq:schemeP}
	\mathrm\ {z}^{a,(n)}(X) = {{W}^{a,(n)} \cdot {f}^{(n-1)}(X) + {b}^{a,(n)}}
\end{eqnarray}
\begin{eqnarray}
\label{eq:schemeP}
	\mathrm\ {f}^{a,(n)}(X) = {\sigma({z}^{a,(n)}(X))}
\end{eqnarray}
\begin{eqnarray}
\label{eq:schemeP}
	\mathrm\ {z}^{a,(n+1)}(X) = {{W}^{a,(n+1)} \cdot {f}^{a,(n)}(X) + {b}^{a,(n+1)}}
\end{eqnarray}
\begin{eqnarray}
\label{eq:schemeP}
	\mathrm\ {f}^{a,(n+1)}(X) = {\Phi({z}^{a,(n+1)}(X))}
\end{eqnarray}

\smallskip
Residual Block,
\begin{eqnarray}
\label{eq:schemeP}
	\mathrm\ {z}^{(n)}(X) = {{W}^{(n)} \cdot {f}^{(n-1)}(X) + {b}^{(n)}}
\end{eqnarray}
\begin{eqnarray}
\label{eq:schemeP}
	\mathrm\ {f}^{(n)}(X) = {\sigma({z}^{(n)}(X))} 
\end{eqnarray}
\begin{eqnarray}
\label{eq:schemeP}
	\mathrm\ {z}^{(n+1)}(X) = {{W}^{(n+1)} \cdot {f}^{(n)}(X) + {b}^{(n+1)}}
\end{eqnarray}
\begin{eqnarray}
\label{eq:schemeP}
	\mathrm\ {z}^{(n+1)}(X) +  {f}^{(n-1)}(X) 
\end{eqnarray}
\begin{eqnarray}
\label{eq:schemeP}
	\mathrm\ {f}^{(n+1)}(X) = {\sigma({z}^{(n+1)}(X) +  {f}^{(n-1)}(X))} 
\end{eqnarray}

\smallskip
Combine Self-Attention Module with Residual Block,
\begin{eqnarray}
\label{eq:schemeP}
	\mathrm\ {{f}^{(n+1)}(X) \otimes{} {f}^{a,(n+1)}(X)} 
\end{eqnarray}

To construct the Attention-ResNet, we stack attention enhanced residual blocks. Assuming the underlying mapping for the plain ANN is ${F}(\mathbf{X},\Theta)$, the residual network asymptotically approximates equation (6), where ${R}(\mathbf{X},\Theta)$ represents the learned residual mapping. In equation (7), attention mask ${M}(\mathbf{X},\Theta)$ is applied via element-wise product to enhance predictive power and model flexibility \cite{he2016deep,wang2017residual}.

\begin{eqnarray}
\label{eq:schemeP}
	\mathrm\ {\mathbf{Y}} =  {R}(\mathbf{X},\Theta) + \mathbf{X}
\end{eqnarray}

\begin{eqnarray}
\label{eq:schemeP}
	\mathrm\ {\mathbf{Y}} = ( {R}(\mathbf{X},\Theta) + \mathbf{X}) \cdot {M}(\mathbf{X},\Theta)
\end{eqnarray}

Since our target variable is binary (see section 4.1), we minimize the error between the estimated conditional probability and the correct target label formulated as the following cross-entropy loss with weight regularization:

\begin{equation}
\begin{split}
  \Theta^{Optimal}=&\argmin_{\Theta}\{\frac{-1}{M}\ \sum_{m} y^{(m)} \cdot \text{log} \mathbf{F}(x^{(m)} ; \Theta) \\
	&+ (1-y^{(m)}) \cdot \text{log} (1 - \mathbf{F}(x^{(m)} ; \Theta)\\
	&+ \lambda \sum_{n} ||\Theta||_{F}^2 \} 
\end{split}
\end{equation}
\smallskip

Cross-entropy loss function speeds up convergence when trained with gradient descent algorithm, also imposes a heavy penalty if $p(y=1|X)=0$ when the true target label is 1, and vice versa. \\

\subsection{3.2 Residual Switching Network (Guided-Attention)}

The switching module is a guided-attention mechanism conditional on SP500 realized volatility, squared VIX, and SP500 variance risk premium. As the market regime shifts from bullish to bearish, the switching module can guide the ResNet to switch from momentum to reversal predictors, and vice versa. Shown in Figure \ref{main}, the switching module (in green, guided by market conditions) computes a conditional weight mask that is then combined with the main module ResNet (in blue) right before the output. More formally: \\

${f}^{(s,0)}(X^{s})=X^{s}$ are market condition feature inputs for the switching module. At layer N, $n=N$, ${f}^{s,(N)}(X^{s})$ computes the conditional mask weights, which normalizes ${z}^{s,(N)}(X^{s})$ using SoftMax funciton, $\Phi$. The conditional weights are normalized to strictly positive and sums to 1.

\begin{eqnarray}
\label{eq:schemeP}
	\mathrm\ {f}^{s,(N)}(X^{s}) = {\Phi({z}^{s,(N)}(X^{s}))}
\end{eqnarray}

${\Phi({z}^{s,(N)}(X^{s}))} = \bigg[ \frac{exp({z}^{s,(N)}_{1}(X^{s}))}{\sum_{c} exp({z}^{s,(N)}_{c}(X^{s}))}, ... , \frac{exp({z}^{s,(N)}_{c}(X^{s}))}{\sum_{c} exp({z}^{s,(N)}_{c}(X^{s}))} \bigg]^\top$
	
\bigskip
Now, we apply conditional weight mask, ${f}^{s,(N)}(X)$, to the Nth parameter layer of the main module, i.e. blue ResNet in Figure \ref{main}.\\
\begin{eqnarray}
\label{eq:schemeP}
	\mathrm\ {f}^{(N)}(X) \otimes{} {f}^{s,(N)}(X^{s})
\end{eqnarray}

\bigskip

The main module in Figure \ref{main} learns both momentum and reversal predictors for approximately 2000 stocks listed in the US. In a bullish regime, the conditional weight mask can place higher weights on momentum predictors and near-zero weights on reversal predictors. As the market enters a bearish regime, the conditional weight mask switches higher weights onto reversal predictors and near-zero weights onto momentum predictors. \\

\begin{eqnarray}
\label{eq:schemeP}
	\mathrm\ {\mathbf{Y}} = ( {R}(\mathbf{X},\Theta) + \mathbf{X}) \cdot {M}(\mathbf{X^{s}},\Theta^{s})
\end{eqnarray}

\subsection{3.3 Training Paradigm}

For the self-attention ResNet, we use 22 ResNet layers for the main module and 12 fully connected layers for the self-attention module. Each parameter layer has dimension 33 to match the dimension of input tensors (momentum and reversal features). For the Switching ResNet, we use 6 ResNet layers for the main module, each parameter layer has dimension 33, and also 6 ResNet layers for the switching module, each parameter layer has dimension 41 to match the dimension of (market condition features). See figure \ref{main}. We use the leaky-ReLU activation function with batch size set to 512. We implement batch normalization \cite{ioffe2015batch} on every hidden layer except the output layer. We first add random Gaussian noise $N(0, 0.1)$ to the input tensor for noise resistance and robustness. We set $\lambda$ to 50 with exponential decay of 0.999, and set dropout rate to 0.5 for every hidden layer. We use the ADAM optimizer with an initial learning rate of 1e-3 and 0.995 exponential decay. We train the proposed networks for approximately 5 epochs and validate our networks every 10k steps.\\

\begin{figure}
\begin{center}
\includegraphics[width=8.5cm]{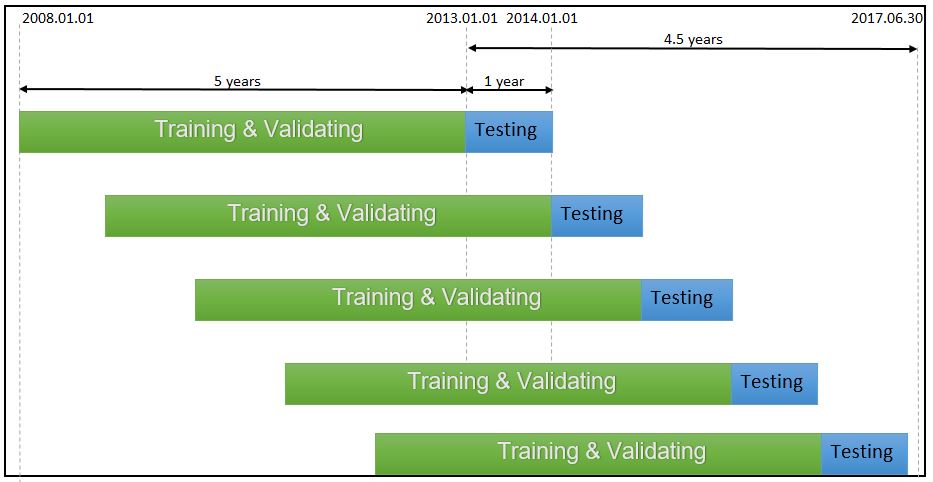}
\end{center}
\caption{Rolling dataset arrangement for training, validating, and testing from Jan 2008 to H12017.}
\label{rolling}
\end{figure}

Figure \ref{rolling} depicts the rolling dataset arrangement for the entire sample period. We employ five years’ of historical data to train and validate our model, 90 percent for training, and 10 percent for validation. We utilize the subsequent one-year's data to test the performance. The procedure continues from Jan 2008 to H12017 to obtain in-sample results from Jan 2008 to the end of 2012, and out-of-sample results from Jan 2013 to H12017.\\

\section{Experiments}
In this section, we carry out experiments to demonstrate the effectiveness of our proposed networks and evaluate their performance. In section 4.1, we describe the dataset in our experiments. Section 4.2 validates the effectiveness of Attention-ResNet. We demonstrate the benefits of the switching module in section 4.3. We further illustrate the performance boost from Switching-ResNet in 4.4.

\subsection{4.1 Dataset} \label{sc_dataset}

The trading universe comprises of approx. 2000 stocks listed on NYSE, NASDAQ, and AMEX. Our sample period spans 9.5 years beginning in Jan 2008 and ending on June 2017. The dataset includes stocks' adjusted daily closing price data and daily VIX index data from Bloomberg. To ensure liquidity, we sample stocks with price traded above 5 USD and market capital over one billion USD at the time of portfolio formation \cite{chen2017slow}. We implement the same model input features as in Takeuchi and Lee 2013. These are price momentum and reversal features, which include 12 normalized monthly returns for month $t-2$ through to $t-13$, 20 normalized past daily returns, and an indicator variable for the month of January to capture turn-of-the-year-patterns. Input features for switching module are 41 market condition features on the SP500, these include 12 normalized SP500 return variance for month $t-2$ through to $t-13$, 23 past daily squared VIX values, and 6 SP500 variance risk premiums for month $t-1$ through $t-6$. For target, we label stocks with monthly return above the median as class 1, and class 0 otherwise. The holding period spans 20 trading days \cite{takeuchi2013applying}. 

\subsection{4.2 Validation of Attention-ResNet}

\noindent{\textbf{Experiment Setting:}} We back-test trading signals through empirical data. We construct delta-neutral long-short portfolio by ranking the estimated conditional probabilities for all stocks in the trading universe, then long and short stocks with estimated probabilities in the top and bottom decile, respectively. We also back-test the predictive accuracy in terms of cross-entropy error. We evaluate and compare the performance of the Attention ResNet with the plain ANN and the linear model (logistic regression).\\

\begin{figure}
  \centering
  \begin{minipage}[b]{0.45\textwidth}
  \centering
    \includegraphics[height=5.5cm, width=8.9cm]{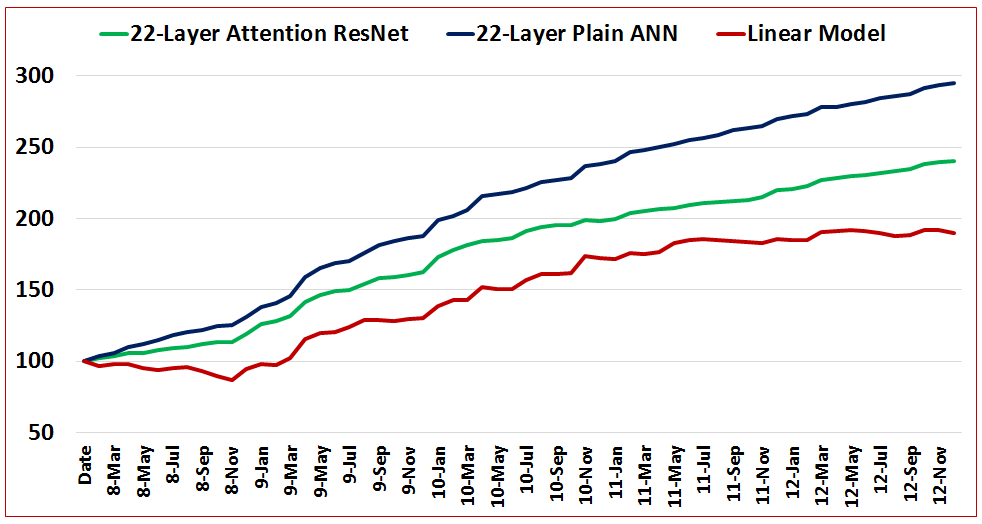}
    \includegraphics[height=5.5cm, width=8.9cm]{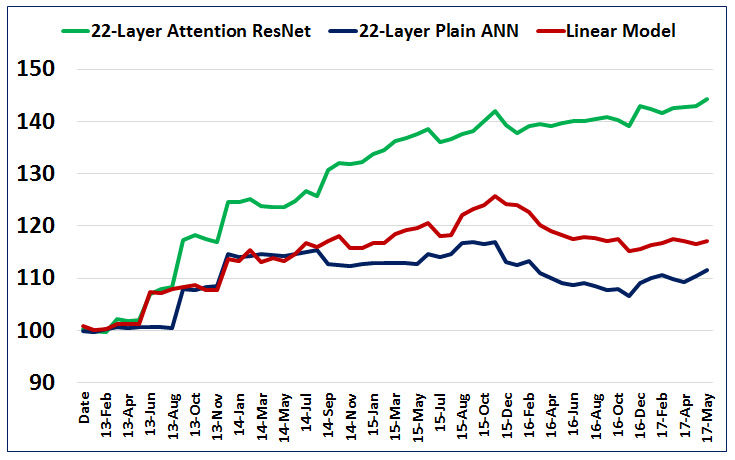}
    \caption{Top: In-sample historical PNL comparison. Bottom: Out-of-sample historical PNL comparison.}
    \label{In-pnl}
  \end{minipage}
\end{figure}

\noindent{\textbf{Results and Discussion:}} The in-sample predictive accuracy (cross-entropy error) of the 22-layer ANN outperformed the 22-layer Attention-ResNet, and the Attention-ResNet outperformed the logistic regression, see Table 1. Similarly, the in-sample historical PNL of the plain ANN outperformed the Attention-ResNet, and the Attention-ResNet outperformed and the linear model, see Fig \ref{In-pnl}. This validates our hypothesis that the 22-layer plain ANN over-fits the dataset and the linear model under-fits the dataset. The 22-layer Attention ResNet can strike an optimal balance between the two. The out-of-sample cross-entropy error of the 22-layer Attention-ResNet outperformed the other two models. \\

\begin{table}[h]
\centering
\small
\begin{tabular}{cccc}
\hline
Year & Attention ResNet & ANN & Linear Model \\
\hline
\hline
In-Sample Mean & 0.6783 & 0.6586 & 0.6918 \\
Out-Sample Mean & 0.6838 & 0.7014 & 0.6931 \\
\hline
\end{tabular}
\caption{Cross-entropy error. In sample 2008 to 2012, out-sample 2013 to H12017.}
\label{tab1}
\end{table} 

The out-of-sample predictive accuracy (cross-entropy error) of the Attention ResNet outperformed the other two models, see Table 1.
The out-of-sample historical PNL of the Attention ResNet also outperformed the other two models, see Fig \ref{In-pnl}. In terms of annualized return, the 22-layer Attention-ResNet averaged 8.86\%, the other two models averaged just over 2\%. Sharpe ratio for the 22-layer Attention-ResNet averaged 1.77, whereas the other two models averaged below 1. We further validate our results by a t-test with the null hypothesis stating no statistically significant difference between the models’ out-of-sample monthly returns. The t-test rejects the null at the 10 percent level and supports the superior out-of-sample performance of Attention-ResNet.\\

\begin{table}[h]
\centering
\small
\begin{tabular}{cccc}
\hline
Year & Attention ResNet & ANN & Linear Model \\
 & (Return) & (Return) & (Return) \\
\hline
\hline
2008 & 18.69\% & 30.74\% & -5.67\% \\
2009 & 43.79\% & 57.24\% & 36.12\% \\
2010 & 37.37\% & 52.25\% & 40.93\% \\
2011 & 20.54\% & 31.71\% & 13.41\% \\
2012 & 24.16\% & 27.45\% & 9.30\% \\
\hline
2013 & 16.93\% & 8.57\% & 7.74\% \\
2014 & 14.91\% & 3.74\% & 8.07\% \\
2015 & 10.27\% & 4.69\% & 9.97\% \\
2016 & -2.95\% & -10.43\% & -10.63\% \\ 
H12017 & 5.16\% & 4.94\% & -1.95\% \\
\hline
\textbf{Out-Sample Mean} & \textbf{8.86\%} & \textbf{2.30\%} & \textbf{2.64\%}\\
\hline
\end{tabular}
\caption{Annualized return. In sample 2008 to 2012, out-sample 2013 to H12017.}
\label{tab1}
\end{table} 

\begin{table}
\centering
\small
\begin{tabular}{cccc}
\hline
Year & Attention ResNet & ANN & Linear Model \\
 & (Sharpe) & (Sharpe) & (Sharpe) \\
\hline
\hline
2008 & 4.96 & 8.99 & -0.69 \\
2009 & 5.98 & 6.64 & 3.68 \\ 
2010 & 4.57 & 5.27 & 3.22 \\ 
2011 & 5.60 & 8.19 & 2.12 \\ 
2012 & 7.70 & 7.36 & 1.43 \\
\hline
2013 & 2.17 & 1.44 & 1.58 \\
2014 & 2.15 & 0.72 & 1.41 \\ 
2015 & 3.31 & 2.14 & 2.51 \\
2016 & -0.98 & -3.09 & -4.43 \\ 
H12017 & 2.18 & 2.84 & 2.40 \\
\hline
\textbf{Out-Sample Mean} & \textbf{1.77} & \textbf{0.81} & \textbf{0.69}\\
\hline
\end{tabular}
\caption{Sharpe ratio. In-sample 2008 to 2012, out-sample 2013 to H12017. Sharpe ratio defined as $\mu$-$r$/$s$, where $\mu$, $r$, $s$ are the annualized return, risk free rate and standard deviation of the PNL.}
\label{tab1}
\end{table}

\subsection{4.3 Validation of Switching Module}\label{sc_seg}

\noindent{\textbf{Experiment Setting:}} We first test the predictive power of the standalone switching module on S\&P500 index's realized volatility (RV). We illustrate that the switching module can accurately predict S\&P500's RV. Our sample period is approx. 13 years beginning in Jan 2005 and ending in Dec 2017. We employ ten years’ of historical data for training and validation, and the subsequent three years to test the out-of-sample $R^{2}$. The target is next month’s S\&P500's RV, $1/22 \sum^{22}_{t=1} {ln ({SP}_{t} / {SP}_{t-1})}^{2}$. There are 41 input features (see section 4.1), the variance risk premium can be quantified as the difference of variance swap rate and ex-post RV in equations below \cite{carr2008variance}, using statistical probability ${\mathbb{P}}$, ${m}_{t,T}= {{M}_{t,T}}/{\mathop{{}\mathbb{E}^{\mathbb{P}}_{t}}[{M}_{t,T}]}$ where ${M}_{t,T}$ is a pricing kernel. We use VIX as an approximator for 30-day variance swap rate on the S\&P500, and formulate the variance risk premium for S\&P500 returns as the difference between VIX and S\&P500 index's one month realized variance \cite{bollerslev2004dynamic}.\\

\noindent{\textbf{Results and Discussion:}}
We evaluate predictive accuracy performance in terms $R^{2}$ for different models: simple strategy, linear regression, 1-3 layer ANN, 2-6 layer ResNet. The simple risk model is often used as the benchmark for out-of-sample $R^{2}$ evaluation \cite{bollerslev2018risk}. Shown in Table \ref{tab4}, the out-of-sample $R^{2}$ improves with 1-layer and 2-layer plain ANNs, but the 3-layer ANN over-fits the dataset with very high in-sample $R^{2}$ and deteriorated out-of-sample $R^{2}$. The ResNet architecture exhibit better control for over-fitting. ResNet's in-sample $R^{2}$ steadily increases with gradual improvement in out-of-sample $R^2$, and we conclude that the 6-layer ResNet demonstrates superior predictive power for S\&P500 realized volatility. \\

\begin{table}
\centering
\small
\begin{tabular}{ccc}
\hline
Network Architecture & In-Sample $R^{2}$   & Out-of-Sample $R^{2}$  \\
\hline
\hline
Simple Strategy & 43\% & 57\%  \\
Linear Regression & 65\% & 58\%  \\
\hline
1-layer ANN & 71\% & 63\%  \\
2-layer ANN & 78\% & 68\%  \\
3-layer ANN & 85\% & 59\%  \\
\hline
2-layer ResNet & 67\% & 56\%  \\
4-layer ResNet & 69\% & 57\%  \\
6-layer ResNet & 73\% & 71\%  \\
\hline
\end{tabular}
\label{tab4}
\caption{$R^{2}$ scores for S\&P500 realized volatility prediction.}
\end{table}

\begin{eqnarray}
\label{eq:schemeP}
	\mathrm\ {SW}_{t,T} = \frac{\mathop{{}\mathbb{E}^{\mathbb{P}}_{t}}[{M}_{t,T} RV_{t,T}]}{\mathop{{}\mathbb{E}^{\mathbb{P}}_{t}}[{M}_{t,T}]} = {\mathop{{}\mathbb{E}^{\mathbb{P}}_{t}}[{m}_{t,T} RV_{t,T}]}
\end{eqnarray}

\begin{eqnarray}
\label{eq:schemeP}
	\mathrm\ {SW}_{t,T} = {\mathop{{}\mathbb{E}^{\mathbb{P}}_{t}}}[{m}_{t,T} RV_{t,T}] = {\mathop{{}\mathbb{E}^{\mathbb{P}}_{t}}[ RV_{t,T}]} + {\mathrm{Cov}^{\mathbb{P}}_{t}}({m}_{t,T},{RV}_{t,T})
\end{eqnarray}
\bigskip

\begin{table}
\centering
\small
\begin{tabular}{ccc}
\hline
Network Architecture & In-Sample $R^{2}$   & Out-of-Sample $R^{2}$  \\
\hline
\hline
Simple Strategy & 43\% & 57\%  \\
Linear Regression & 65\% & 58\%  \\
\hline
1-layer ANN & 71\% & 63\%  \\
2-layer ANN & 78\% & 68\%  \\
3-layer ANN & 85\% & 59\%  \\
\hline
2-layer ResNet & 67\% & 56\%  \\
4-layer ResNet & 69\% & 57\%  \\
6-layer ResNet & 73\% & 71\%  \\
\hline
\end{tabular}
\label{tab4}
\caption{$R^{2}$ scores for S\&P500 realized volatility prediction.}
\end{table}

\begin{table}[h]
\centering
\small
\begin{tabular}{ccc}
\hline
Year & Attention ResNet  &  Switching ResNet \\
 & (Self Attention) &  (Guided Attention)  \\
\hline
\hline
 & Annual Return &  Annual Return \\
\hline
2013 & 16.93\% & 12.20\% \\
2014 & 14.91\% & 13.41\% \\
2015 & 10.27\% & 14.38\% \\
2016 & -2.95\% & 8.30\% \\
H1 2017 & 5.16\% & 5.49\% \\
\textbf{Mean} & \textbf{8.86\%} & \textbf{10.76\%} \\
\hline
\hline
 & Sharpe Ratio &  Sharpe Ratio \\
\hline
2013 & 2.17 & 2.08 \\
2014 & 2.15  & 1.93  \\
2015 & 3.31 & 2.43  \\
2016 & -0.98  & 2.52  \\
H1 2017 & 2.18  & 2.15  \\
\textbf{Mean} & \textbf{1.77} & \textbf{2.22} \\
\hline
\end{tabular}
\label{tab5}
\caption{Out-of-sample return and Sharpe ratio.}
\end{table}

\subsection{4.4 Validation of Switching-ResNet}

\noindent{\textbf{Experiment Setting:}} We combine the switching module (6-layer ResNet from section 4.2) with the main module to complete the Switching-ResNet, as described in section 3.2. We evaluate the ability of the switching module to switch between momentum and reversal predictors as market regimes changes (bullish or bearish). We analyze the dynamic behavior of the conditional weight mask. Finally, we evaluate the out-of-sample performance of the Switching-ResNet. \\

\noindent{\textbf{Results and Discussion:}} The dynamic behavior of the conditional weight mask demonstrates the effectiveness guided-attention on momentum and reversal predictors as market conditions changes from bullish to bearish market regime, and vice versa. The switching module computes 33 weights in the range [0,1] and the sum of the weights is 1. We observe two patterns on the weight mask. Weights assigned to reversal predictors are higher during periods of high market volatility and positively correlated with the VIX index. Weights assigned to momentum predictors are lower during periods of high market volatility and negatively correlated with the VIX index. \\

\begin{figure}
  \centering
  \begin{minipage}[b]{0.45\textwidth}
  \centering
    \includegraphics[height=3cm, width=9cm]{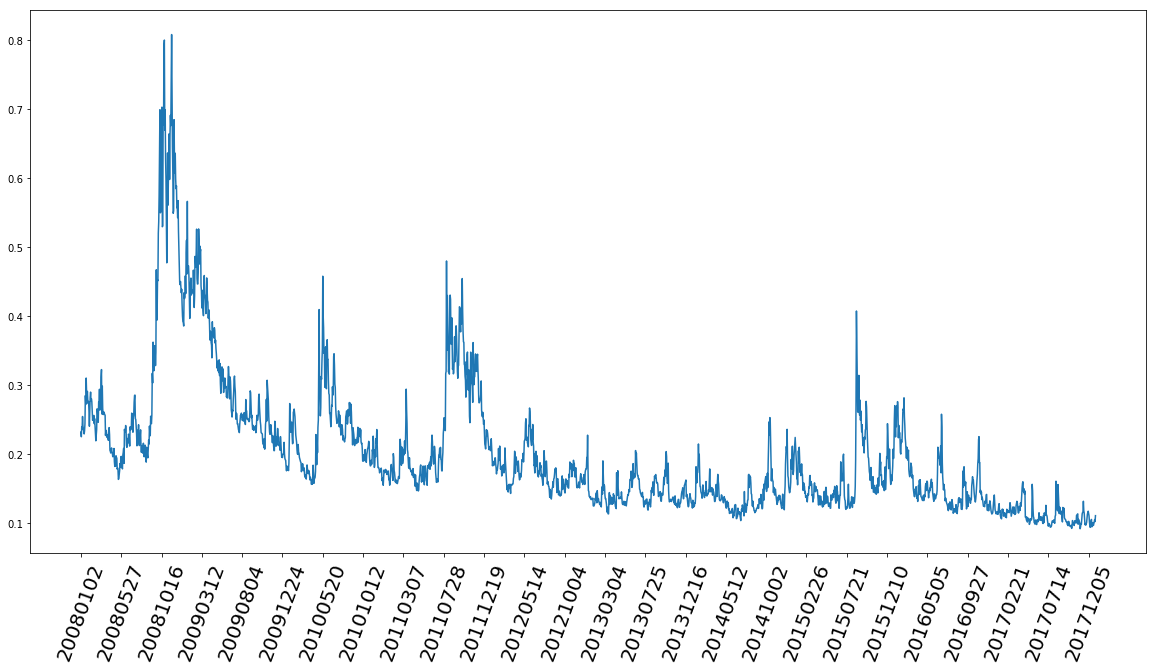}
  \end{minipage}
  \begin{minipage}[b]{0.45\textwidth}
  \centering
    \includegraphics[height=3cm, width=9cm]{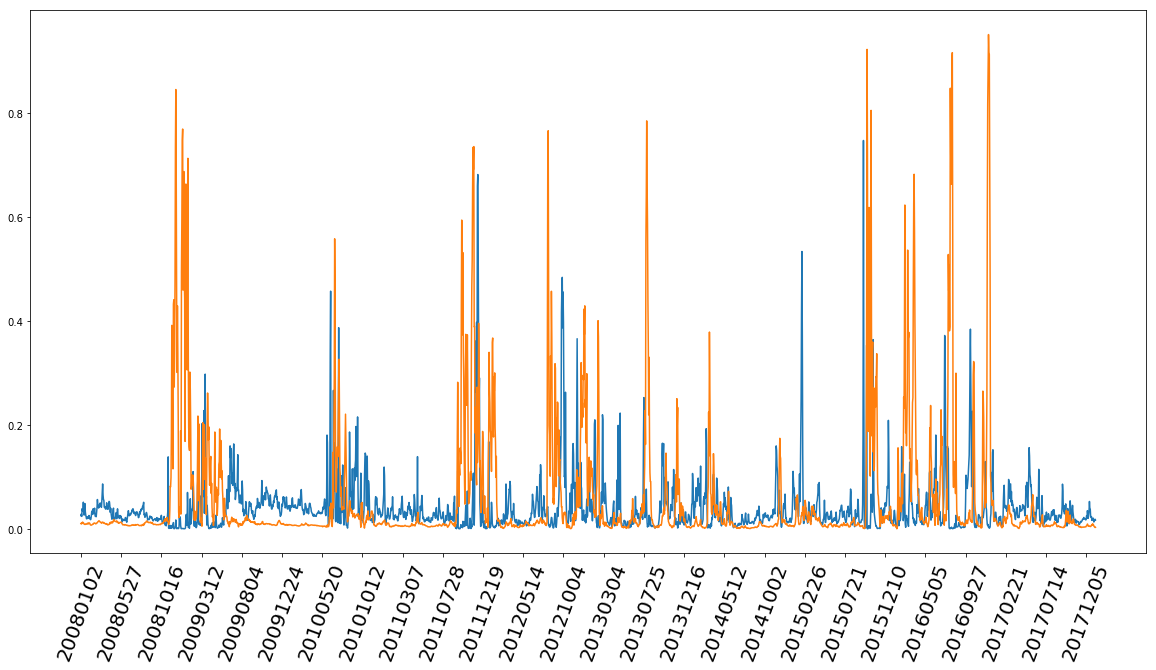}
  \end{minipage}
  \begin{minipage}[b]{0.45\textwidth}
  \centering
    \includegraphics[height=3cm, width=9cm]{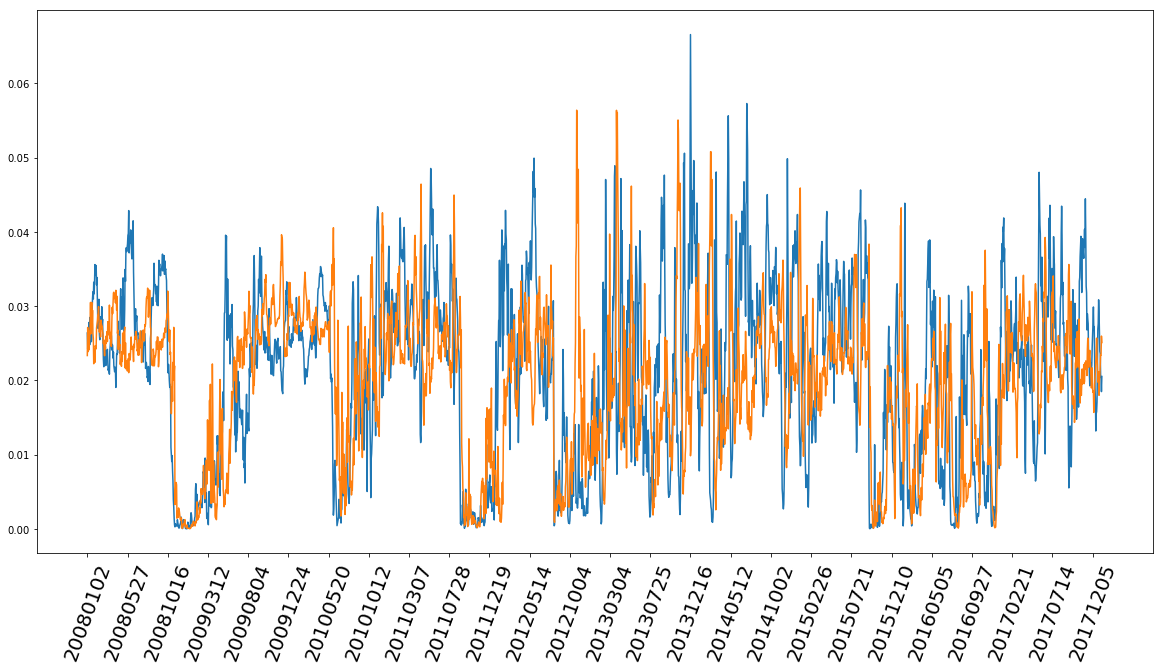}
    \caption{Top:  VIX index from Jan 2008 to Dec 2017. Middle: Weights assigned to reversal predictors.  Bottom: Weights assigned to momentum predictors.}
    \label{weights}
  \end{minipage}
\end{figure}

Figure \ref{weights} plots the co-movement of VIX index and conditional weights. When VIX spiked in Q3 2008, Q2 2010, Q3 2011 and H2 2015, weights on reversal predictors spiked to 0.8, and weights assigned to momentum predictors dipped. The correlation matrix below shows that the VIX index and reversal weights are correlated by 23\%, whereas the VIX index and momentum weights are correlated by -19\%. The conditional weights on reversal and momentum predictors are correlated by -59\%. This result further supports our hypothesis that in a bullish regime with lower VIX, the conditional weight mask gives more attention to momentum predictors, and as the market changes to bearish regime with higher VIX, the switching model guides attention back on reversal predictors. 
\smallskip

\begin{table}[h]
\centering
\small
\begin{tabular}{|c|ccc| }
\hline
 & VIX Index & Reversal & Momentum  \\
 &   & Weight & Weight \\
\hline
VIX Index      & 100\% & 23\%  & -19\% \\
Reversal Weight &  & 100\% & -59\%\\
Momentum Weight &  &       & 100\%\\
\hline
\end{tabular}
\label{tab4}
\end{table} 
\smallskip

The Switching-ResNet's out-of-sample historical PNL outperformed the Attention-ResNet in 2016, and their performance for other years was relatively similar, see Table 5. When the S\&P500 volatility and the VIX index spiked in late 2015 to mid-2016, the Attention-ResNet was not profitable during this period. The Switching-ResNet was able to correctly identify the change in market conditions, and guide attention from momentum predictors back onto reversal predictors to reap higher rewards in the US equities market.\\

\section{Conclusion}

We propose the Residual Switching Network to combine a regime-switching module with the residual network. The advantages of our network are in two folds. The first advantage comes from leveraging the guided-attention mechanism to automatically sense dynamic changes in stock market conditions, and guide our network to place more attention on momentum and reversal predictors accordingly. Our experiment validates the effectiveness of the switching module and provides direct evidence for our network's ability to generate superior out-of-sample Sharpe ratio as market conditions change over time. The second advantage lies with residual learning's ability to train very deep networks while addressing the over-fitting problem when applying DL models to noisy financial data. Our experiments verify that our deep Attention-ResNet can strike a balance between linear and complex non-linear models for financial modeling. The scope of our work extends residual learning and attention mechanism to portfolio management, these concepts are hopeful for other financial fields.

\clearpage

\begin{rezabib}

@article{ioffe2015batch,
  title={Batch normalization: Accelerating deep network training by reducing internal covariate shift},
  author={Ioffe, Sergey and Szegedy, Christian},
  journal={arXiv preprint arXiv:1502.03167},
  year={2015}
}

@article{avellaneda2010statistical,
  title={Statistical arbitrage in the US equities market},
  author={Avellaneda, Marco and Lee, Jeong-Hyun},
  journal={Quantitative Finance},
  volume={10},
  number={7},
  pages={761--782},
  year={2010},
  publisher={Taylor \& Francis}
}

@article{dixon2018sequence,
  title={Sequence classification of the limit order book using recurrent neural networks},
  author={Dixon, Matthew},
  journal={Journal of computational science},
  volume={24},
  pages={277--286},
  year={2018},
  publisher={Elsevier}
}

@article{cherkassky2004practical,
  title={Practical selection of SVM parameters and noise estimation for SVM regression},
  author={Cherkassky, Vladimir and Ma, Yunqian},
  journal={Neural networks},
  volume={17},
  number={1},
  pages={113--126},
  year={2004},
  publisher={Elsevier}
}

@incollection{takeuchi2013applying,
  title={Applying deep learning to enhance momentum trading strategies in stocks},
  author={Takeuchi, Lawrence and Lee, Yu-Ying Albert},
  booktitle={Technical Report},
  year={2013},
  publisher={Stanford University}
}

@article{das2012support,
  title={Support vector machines for prediction of futures prices in Indian stock market},
  author={Das, Shom Prasad and Padhy, Sudarsan},
  journal={International Journal of Computer Applications},
  volume={41},
  number={3},
  year={2012},
  publisher={Citeseer}
}

@book{gately1995neural,
  title={Neural networks for financial forecasting},
  author={Gately, Edward},
  year={1995},
  publisher={John Wiley \& Sons, Inc.}
}

@article{huang2005forecasting,
  title={Forecasting stock market movement direction with support vector machine},
  author={Huang, Wei and Nakamori, Yoshiteru and Wang, Shou-Yang},
  journal={Computers \& Operations Research},
  volume={32},
  number={10},
  pages={2513--2522},
  year={2005},
  publisher={Elsevier}
}

@inproceedings{he2016deep,
  title={Deep residual learning for image recognition},
  author={He, Kaiming and Zhang, Xiangyu and Ren, Shaoqing and Sun, Jian},
  booktitle={Proceedings of the IEEE conference on computer vision and pattern recognition},
  pages={770--778},
  year={2016}
}

@inproceedings{wang2017residual,
  title={Residual attention network for image classification},
  author={Wang, Fei and Jiang, Mengqing and Qian, Chen and Yang, Shuo and Li, Cheng and Zhang, Honggang and Wang, Xiaogang and Tang, Xiaoou},
  booktitle={Proceedings of the IEEE Conference on Computer Vision and Pattern Recognition},
  pages={3156--3164},
  year={2017}
}

@article{deng2017deep,
  title={Deep direct reinforcement learning for financial signal representation and trading},
  author={Deng, Yue and Bao, Feng and Kong, Youyong and Ren, Zhiquan and Dai, Qionghai},
  journal={IEEE transactions on neural networks and learning systems},
  volume={28},
  number={3},
  pages={653--664},
  year={2017},
  publisher={IEEE}
}

@article{dixon2018high,
  title={A high-frequency trade execution model for supervised learning},
  author={Dixon, Matthew},
  journal={High Frequency},
  volume={1},
  number={1},
  pages={32--52},
  year={2018},
  publisher={Wiley Online Library}
}

@article{bao2017deep,
  title={A deep learning framework for financial time series using stacked autoencoders and long-short term memory},
  author={Bao, Wei and Yue, Jun and Rao, Yulei},
  journal={PloS one},
  volume={12},
  number={7},
  pages={e0180944},
  year={2017},
  publisher={Public Library of Science}
}

@article{hinton2006reducing,
  title={Reducing the dimensionality of data with neural networks},
  author={Hinton, Geoffrey E and Salakhutdinov, Ruslan R},
  journal={science},
  volume={313},
  number={5786},
  pages={504--507},
  year={2006},
  publisher={American Association for the Advancement of Science}
}

@article{simonyan2014very,
  title={Very deep convolutional networks for large-scale image recognition},
  author={Simonyan, Karen and Zisserman, Andrew},
  journal={arXiv preprint arXiv:1409.1556},
  year={2014}
}

@inproceedings{szegedy2015going,
  title={Going deeper with convolutions},
  author={Szegedy, Christian and Liu, Wei and Jia, Yangqing and Sermanet, Pierre and Reed, Scott and Anguelov, Dragomir and Erhan, Dumitru and Vanhoucke, Vincent and Rabinovich, Andrew},
  booktitle={Proceedings of the IEEE conference on computer vision and pattern recognition},
  pages={1--9},
  year={2015}
}

@article{hinton2012deep,
  title={Deep neural networks for acoustic modeling in speech recognition: The shared views of four research groups},
  author={Hinton, Geoffrey and Deng, Li and Yu, Dong and Dahl, George E and Mohamed, Abdel-rahman and Jaitly, Navdeep and Senior, Andrew and Vanhoucke, Vincent and Nguyen, Patrick and Sainath, Tara N and others},
  journal={IEEE Signal processing magazine},
  volume={29},
  number={6},
  pages={82--97},
  year={2012},
  publisher={IEEE}
}

@article{kaastra1996designing,
  title={Designing a neural network for forecasting financial and economic time series},
  author={Kaastra, Iebeling and Boyd, Milton},
  journal={Neurocomputing},
  volume={10},
  number={3},
  pages={215--236},
  year={1996},
  publisher={Elsevier}
}

@article{saad1998comparative,
  title={Comparative study of stock trend prediction using time delay, recurrent and probabilistic neural networks},
  author={Saad, Emad W and Prokhorov, Danil V and Wunsch, Donald C},
  journal={IEEE Transactions on neural networks},
  volume={9},
  number={6},
  pages={1456--1470},
  year={1998},
  publisher={IEEE}
}

@inproceedings{dixon2015implementing,
  title={Implementing deep neural networks for financial market prediction on the Intel Xeon Phi},
  author={Dixon, Matthew and Klabjan, Diego and Bang, Jin Hoon},
  booktitle={Proceedings of the 8th Workshop on High Performance Computational Finance},
  pages={6},
  year={2015},
  organization={ACM}
}

@article{itti2001computational,
  title={Computational modelling of visual attention},
  author={Itti, Laurent and Koch, Christof},
  journal={Nature reviews neuroscience},
  volume={2},
  number={3},
  pages={194},
  year={2001},
  publisher={Nature Publishing Group}
}

@inproceedings{walther2002attentional,
  title={Attentional selection for object recognition—a gentle way},
  author={Walther, Dirk and Itti, Laurent and Riesenhuber, Maximilian and Poggio, Tomaso and Koch, Christof},
  booktitle={International Workshop on Biologically Motivated Computer Vision},
  pages={472--479},
  year={2002},
  organization={Springer}
}

@article{kingma2014adam,
  title={Adam: A method for stochastic optimization},
  author={Kingma, Diederik P and Ba, Jimmy},
  journal={arXiv preprint arXiv:1412.6980},
  year={2014}
}

@article{chen2017slow,
  title={Slow diffusion of information and price momentum in stocks: Evidence from options markets},
  author={Chen, Zhuo and Lu, Andrea},
  journal={Journal of Banking \& Finance},
  volume={75},
  pages={98--108},
  year={2017},
  publisher={Elsevier}
}

@article{jegadeesh1993returns,
  title={Returns to buying winners and selling losers: Implications for stock market efficiency},
  author={Jegadeesh, Narasimhan and Titman, Sheridan},
  journal={The Journal of finance},
  volume={48},
  number={1},
  pages={65--91},
  year={1993},
  publisher={Wiley Online Library}
}

@article{vayanos2013institutional,
  title={An institutional theory of momentum and reversal},
  author={Vayanos, Dimitri and Woolley, Paul},
  journal={The Review of Financial Studies},
  volume={26},
  number={5},
  pages={1087--1145},
  year={2013},
  publisher={Oxford University Press}
}

@article{schwert1989does,
  title={Why does stock market volatility change over time?},
  author={Schwert, G William},
  journal={The journal of finance},
  volume={44},
  number={5},
  pages={1115--1153},
  year={1989},
  publisher={Wiley Online Library}
}

@article{bollerslev2018risk,
  title={Risk everywhere: Modeling and managing volatility},
  author={Bollerslev, Tim and Hood, Benjamin and Huss, John and Pedersen, Lasse Heje},
  journal={The Review of Financial Studies},
  volume={31},
  number={7},
  pages={2729--2773},
  year={2018},
  publisher={Oxford University Press}
}

@article{carr2008variance,
  title={Variance risk premiums},
  author={Carr, Peter and Wu, Liuren},
  journal={The Review of Financial Studies},
  volume={22},
  number={3},
  pages={1311--1341},
  year={2008},
  publisher={Oxford University Press}
}

@article{bollerslev2004dynamic,
  title={Dynamic Estimation of Volatility Risk and Investor Risk Aversion from Option Implied and Realized Volatilities},
  author={Bollerslev, T and Gibson, M and Zhou, H},
  journal={Finance and Economics Discussion Working Paper, Federal Reserve Board},
  number={2004-56},
  year={2004}
}

@article{hamilton1989new,
  title={A new approach to the economic analysis of nonstationary time series and the business cycle},
  author={Hamilton, James D},
  journal={Econometrica: Journal of the Econometric Society},
  pages={357--384},
  year={1989},
  publisher={JSTOR}
}

@article{engel1994can,
  title={Can the Markov switching model forecast exchange rates?},
  author={Engel, Charles},
  journal={Journal of International Economics},
  volume={36},
  number={1-2},
  pages={151--165},
  year={1994},
  publisher={Elsevier}
}

@article{kim1998business,
  title={Business cycle turning points, a new coincident index, and tests of duration dependence based on a dynamic factor model with regime switching},
  author={Kim, Chang-Jin and Nelson, Charles R},
  journal={Review of Economics and Statistics},
  volume={80},
  number={2},
  pages={188--201},
  year={1998},
  publisher={MIT Press}
}

@article{bahdanau2014neural,
  title={Neural machine translation by jointly learning to align and translate},
  author={Bahdanau, Dzmitry and Cho, Kyunghyun and Bengio, Yoshua},
  journal={arXiv preprint arXiv:1409.0473},
  year={2014}
}

\end{rezabib}

\small
\bibliographystyle{unsrt}

\end{document}